\documentclass[11pt]{article}

\usepackage{latexsym,amsfonts,amsmath,amssymb,pstricks,wasysym,stmaryrd,enumerate,epsfig}
\usepackage{mathrsfs}
\usepackage{theorem}

\numberwithin{equation}{section}
\newtheorem{definition}{Definition}[section]

\newtheorem{proposition}[definition]{Proposition}

\newtheorem{lemma}[definition]{Lemma}

\newcommand{\be}{\begin{equation}}
\newcommand{\ee}{\end{equation}}
\newcommand{\beu}{\begin{equation*}}
\newcommand{\eeu}{\end{equation*}}
\newcommand{\bea}{\begin{eqnarray}}
\newcommand{\eea}{\end{eqnarray}}
\newcommand{\beaa}{\begin{eqnarray*}}
\newcommand{\eeaa}{\end{eqnarray*}}
\newcommand{\bmx}{\begin{pmatrix}}
\newcommand{\emx}{\end{pmatrix}}

\newcommand{\del}{\partial}

\newcommand{\proof}{{\noindent\textbf{Proof. }}}
\newcommand{\finproof}{{\hfill \rule{5pt}{5pt}}}
\newcommand{\HL}{{\mathcal H}}

\newcommand{\half}{\frac{1}{2}}

\newcommand{\nn}{\nonumber}

\newcommand{\8}{{\infty}}

\newcommand{\liealg}[1]{{\mathfrak{#1}}}

\newcommand{\eps}{\epsilon}
\newcommand{\tr}{\,{\rm tr}}

\newcommand{\Z}{{\mathbb Z}}\newcommand{\CC}{{\mathbb C}}
\newcommand{\Zn}{\mathbb Z_n}
\newcommand{\PP}{\mathbb{P}}
\renewcommand{\P}{{\mathscr P}}
\newcommand{\Q}{{\mathscr Q}}
\newcommand{\R}{{\mathbb R}}
\newcommand{\QQ}{\mathbb{Q}}
\newcommand{\cT}{{\cal{T}}}
\newcommand{\cH}{{\cal{H}}}
\advance\textheight by \topskip
\begin{document}

\baselineskip 17pt
\parindent 18pt
\parskip 9pt

\begin{flushright}
\break

\end{flushright}
\vspace{2cm}
\begin{center}
{\Large {\bf Integrable Models From Twisted Half Loop Algebras}}
{\Large {\bf }}\\[4mm]
\vspace{2cm}
{\large N. Cramp\'e\footnote{crampe@sissa.it} and C. A. S. Young\footnote{charlesyoung@cantab.net}}
\\
{\em Department of Mathematics, University of York,\\
Heslington Lane, York YO10 5DD, UK}

\end{center}

\vskip 1.15in
 \centerline{\small\bf ABSTRACT}
\centerline{
\parbox[t]{5in}{\small
\noindent This paper is devoted to the construction of new integrable quantum-mechanical
models based on certain subalgebras of the half loop algebra of $\liealg{gl}_N$. Various results about these subalgebras are proven by presenting them in the notation of the St Petersburg school. These results are then used to demonstrate the integrability, and find the symmetries, of two types of physical system: twisted Gaudin magnets, and Calogero-type models of particles on several half-lines meeting at a point.}}

\vspace{1cm}

\newpage
\section{Introduction}
This paper has two motivations. On the one hand, we are interested in physical models of particles on a number of half-lines joined at a central point. Such systems, for free particles, have been treated in, for example, \cite{ExnerSeba,Kostrykin}. Here we would like to consider interacting models, to establish that integrable examples of such models exist, and to find their symmetries.
We shall work out explicitly two examples: the Gaudin model \cite{gaudin} and the Calogero model \cite{Cal}. Both have numerous applications 
in physics and in mathematics. For example, the reduced BCS model for conventionnal superconductivity can be diagonalized 
in an algebraic way \cite{ric} using the Gaudin model. Other, more recent, applications of the 
Gaudin model in quantum many-body physics can be found for example in the reviews \cite{lin,duk}. 
Besides being of intrinsic interest due to its exact solvability, the Calogero model plays a role in the study of two dimensional Yang-Mills theory \cite{gor}, the quantum Hall effect \cite{azu} and fractional statistics \cite{pol}.  

Our second motivation is algebraic. The notation of the St Peterburg school \cite{FRT} is a powerful tool when working with the Yangian \cite{Dri} of $\liealg{gl}_N$ and its subalgebras: the reflection algebras \cite{cherednik,sklyanin} and twisted Yangians \cite{Ytwist}. These are quantum algebras, but the construction has a classical limit in which the quantum $R$-matrix and Yang-Baxter Equation are replaced by their classical counterparts (see for example \cite{FT}). The classical limit of the Yangian is the half loop algebra, and the limits of the reflection algebras and twisted Yangian are subalgebras of this half loop algebra defined by automorphisms of order 2. But there also exist, at least in the classical case, other subalgebras of the half loop algebra, defined by automorphisms of higher finite order. We wish to study these subalgebras using classical $r$-matrix techniques.

It is well-known that the half-loop algebras associated to Lie algebras are crucial 
in the study of Gaudin models and Calogero models. 
These algebras provide, in the former case, a systematic way to construct the model 
(see e.g \cite{jur}) and, in the latter case, the symmetry algebras of the system 
\cite{BGHP,hik,cra}. In both cases, they allow one to prove the integrability of the model
.
We shall find similar connections in the cases studied in this paper. Indeed, 
we shall see below that the order $n$ subalgebras of the half loop algebra appears naturally in the description of models on $n$ half-lines.

This paper is structured as follows. We begin with a brief review of
the half loop algebra  of $\liealg{gl}_N$ and its subalgebras
associated to automorphisms of order $n$. We make use of the
notation of the St Petersburg school to find Abelian subalgebras. In
the subsequent sections these algebraic results are shown to provide
new quantum integrable models and demonstrate their symmetries:
section \ref{gaudin} discusses ``twisted'' Gaudin magnets, and
section \ref{calogero} introduces Calogero-type models on $n$
half-lines joined a central point. We end with some conclusions and
a short discussion of classical counterparts of these results.

\section{Half loop algebra and subalgebras}
\subsection{St Petersburg notation and half loop algebra}

The half loop algebra $\HL_N$ based on $\liealg{gl}_N$ is the
complex associative unital algebra with the following set of
generators $\left\{t_{ij}^{(\alpha)}\ \big|\ 1\leqslant i,j
\leqslant N,\alpha\in \mathbb Z_{\geqslant 0}\right\}$, subject to
the defining relations \be [t_{ij}^{(\alpha)},t_{kl}^{(\beta)}]=
\delta_{jk}\ t_{il}^{(\alpha+\beta)}\ -\ \delta_{il}\
t_{kj}^{(\alpha+\beta)} \label{HNalg} \ee for
$\alpha,\beta\geqslant0$ and $1\leqslant i,j,k,l \leqslant N$.
It is isomorphic to the algebra $\liealg{gl}_N[z]$ of
polynomials in an indeterminate $z$ with coefficients in
$\liealg{gl}_N$, with the generators identified as follows: \be
t_{ij}^{(\alpha)}=e_{ij}z^{\alpha},\label{ident}\ee where $e_{ij}$
are the generators of $\liealg{gl}_N$, satisfying the commutation
relations \be\label{lie} \left[ e_{ij},e_{kl} \right] = \delta_{jk}
e_{il} - \delta_{il} e_{kj}.\ee

It will simplify our computations to introduce the notation of the
St Petersburg school: let $E_{ij}$ be the $N\times N$ matrix with a
$1$ in the $ij$th slot and zeros elsewhere. These are the generators
of $\liealg{gl}_N$ in the fundamental representation. Let us now
gather the generators of $\HL_N$ in the matrix
\begin{eqnarray}
\label{def:T} T(u) =\sum_{i,j=1}^{N} E_{ij}\otimes \sum_{\alpha \geqslant 0}
\frac{t_{ji}^{(\alpha)}}{u^{\alpha+1}} =\sum_{i,j=1}^{N} E_{ij}\otimes T_{ji}(u)= \sum_{\alpha \geqslant 0} \frac{T^{(\alpha)}}{u^{\alpha+1}} \;,
\end{eqnarray}
where $T^{(\alpha)}=\sum_{i,j=1}^{N}E_{ij}\otimes t_{ji}^{(\alpha)}$ ($\alpha\geqslant 0$) and $u$ is a formal parameter called the spectral parameter.
Note the flip of the indices between $E_{ij}$ and $t_{ji}$, which will prove convenient later. The algebraic 
object $T(u)$ is an element of $\mathrm{Mat}_{N\times N} \otimes {\HL}_N[[u^{-1}]]$, and
as usual we refer to $\mathrm{Mat}_{N\times N}$ as the auxiliary space and ${\HL}_N$ as the
algebraic space. In what follows we shall require several copies of both spaces. We use letter $a,b,\dots$ from the start of the alphabet to refer to copies of the auxiliary space and numerals $1,2,\dots$ for copies of the algebraic space.
Let us introduce also 
\be
r_{ab}(u)=\frac{P_{ab}}{u}\label{classR}
\ee 
where
$P_{ab}=\sum_{i,j=1}^{N} E_{ij}\otimes E_{ji}$ is the permutation operator between two auxiliary spaces: 
the letters $a$ and $b$ stand respectively for the first and the second spaces. By definition, it satisfies
$P_{ab}\, v\otimes w=w\otimes v$ ($v,w\in \CC^N$). The matrix $r_{ab}(u)$,
usually called the classical R-matrix (see for example \cite{FT}), satisfies the classical
Yang-Baxter equation \be
[r_{ab}(u_a-u_b),r_{ac}(u_a-u_b)]+[r_{ab}(u_a-u_b),r_{bc}(u_b-u_c)]+
[r_{ac}(u_a-u_c),r_{bc}(u_b-u_c)]=0 \, \ee
and allows us to encode the half loop algebra defining relations (\ref{HNalg}) in the
simple equation
\begin{eqnarray}
\left[ T_a(u), T_b(v) \right] = \left[ T_a(u) + T_b(v),
r_{ab}(u-v)\right]\ .\label{halfloop}
\end{eqnarray}
This form of commutation relations can be obtained easily
by taking the classical limit of the presentation of the
Yangian of $\liealg{gl}_N$ \cite{Dri} introduced by 
L.D.~Faddeev, N.Yu.~Reshetikhin and L.A.~Takhtajan of St Petersburg \cite{FRT}. By taking the trace in the space $a$ in (\ref{halfloop}), it is straightforward to show that the coefficients of the series $t(u)=\tr_a T_a(u)$ are central. The quotient of the algebra ${\HL}_N$ by the relation $t(u)=0$ is isomorphic to the polynomial algebra $\liealg{sl}_N[z]$.

The identification (\ref{ident}) between the generators of $\HL_N$ and $\liealg{gl}_N[z]$ now reads
\be 
T_a(u)=\frac{\PP_{a1}}{u-z}\ ,\label{real}
\ee
where
\be\PP_{a1}=\sum_{i,j=1}^{N} E_{ij}\otimes e_{ji}\ee 
and $\frac{1}{u-z}$ is to be understood as the formal series
$\sum_{\alpha\geqslant 0}\frac{z^{\alpha}}{u^{\alpha+1}}$. Note the similarity
between relations (\ref{classR}) and (\ref{real}): the only differences are that the second auxiliary space (denoted $b$) in (\ref{classR}) is replaced by an algebraic space (denoted $1$) and that the spectral parameter is shifted by $z$. In fact, there exists a more general solution of the relations
(\ref{halfloop}) in the $L$-fold tensor product of $\liealg{gl}_N[z]$,
\begin{eqnarray}
T_a(u)=\sum_{\ell=1}^L\frac{\PP_{a\ell}}{u-z_\ell}\ .\label{realL}
\end{eqnarray}
{}From now on, we work in the enveloping algebra $\mathcal U ({\cal
H}_N)$ in which, for example, the product $T_a(u)^2$ makes sense.

\subsection{The inner-twisted algebras}

Let $\sigma$ be an inner automorphism of $\liealg{gl}_N$ of order
$n$. One way to define $\sigma$ is by its action on matrices $X\in
{\rm Mat}_{N\times N}$ in the fundamental representation: \be
\sigma: X \mapsto G^{-1} XG \label{aa}\ee where $G\in {\rm
Mat}_{N\times N}$ satisfies $G^n=1$; the action of $\sigma$ on the
abstract algebra $\liealg{gl}_N$ is then given by $\sigma: e_{ij}
\mapsto G_{jq} (G^{-1})_{pi} e_{pq}$, or, in the notation of the
previous section, \be \sigma: \PP_{a1} \mapsto G_a \PP_{a1}
G_a^{-1}.\label{aaa}\ee

The eigenvalues of $\sigma$ are the $n$-th roots of unity $\tau^k:=\exp 2\pi i k/n$, and for each $k\in\Zn:=\frac{\mathbb Z}{n
\mathbb Z}$ the map \be P_k = \frac{1}{n} \sum_{j\in\Zn} \tau^{-jk}
\sigma^j\ee is the projector onto the $\tau^k$-eigenspace: \be
\sigma P_k = \tau^k P_k,\quad  P_k P_j = \delta_{jk} P_k.\ee Since
$1=P_0 + P_1 +\dots+ P_{n-1}$, $\liealg{gl}_N$ decomposes into the
direct sum of eigenspaces of $\sigma$. This decomposition respects
the Lie bracket, in the sense that if $\sigma e = \tau^k e$ and
$\sigma f = \tau^l f$ then \be \sigma [e,f] = \left[ \sigma e,
\sigma f\right] = \tau^{k+l} [e,f],\ee and is said to be a
$\Zn$-gradation of $\liealg{gl}_N$.

By a change of basis we can take \be G=
\mathrm{diag}(\underbrace{1,\dots, 1}_{N_0},
\underbrace{\tau,\dots,\tau}_{N_1}, \dots,
\underbrace{\tau^{n-1},\dots,\tau^{n-1}}_{N_{n-1}})    \ee
where $N_0 + N_1 +\dots+N_{n-1} = N$. Note that the $+1$-eigenspace
of $\sigma$ is the Lie subalgebra $\liealg{gl}_{N_0} \oplus
\liealg{gl}_{N_1} \oplus\dots\oplus \liealg{gl}_{N_{n-1}}$.

Let us define \be \liealg{gl}_N[z]^\sigma = \left\{ A(z)\in
\liealg{gl}_N[z]\ \big|\ \sigma A(z) = A(\tau z)\right\};\ee that
is, $\liealg{gl}_N[z]^\sigma$ is the subalgebra of
$\liealg{gl}_N[z]$ in which each element of degree $k$ is also in
the $\tau^k$-eigenspace of $\sigma$. There is a surjective
projection map $\liealg{gl}_N[z]\rightarrow\liealg{gl}_N[z]^\sigma$,
defined by $e z^k \mapsto P_k e z^k$. In view of (\ref{aaa}), this
sends
\be T(u) \mapsto \sum_{j\in \Zn} \tau^j G^j T(u\tau^j) G^{-j}=:B(u),\label{Bu}\ee
which defines the formal series $B(u)$
whose expansion
\be B(u)= \frac{1}{u} B^{(0)} + \frac{1}{u^2} B^{(1)} + \dots,\ee
contains by construction a complete set of generators of $\liealg{gl}_N[z]^\sigma$.

\begin{lemma} $B(u)$ obeys
\be\left[ B_a(u), B_b(v) \right] = \sum_{k\in \Zn} \left[ \tau^k
B_a(u) + B_b(v), \frac{G_a^{-k} P_{ab}
G_a^{k}}{u-\tau^kv}\right]\label{Bbrac}\ee and has the property that
for all $k\in \Zn$ \be B(u) = \tau^k G^k B(u\tau^k) G^{-k}
\label{unitarity}.\ee
\end{lemma}
\proof These first of these is true by virtue of (\ref{halfloop}) while the
second follows immediately from the definition (\ref{Bu}). \finproof

The coefficients in the expansion of $b(u) = \tr B(u)$ are central in $\liealg{gl}_N[z]^\sigma$, as may be seen by taking the trace in space $a$ or $b$ in (\ref{Bbrac}). But there exist also other abelian subalgebras in $\mathcal U (\liealg{gl}_N[z]^\sigma)$, as follows.

\begin{proposition}\label{b'b'0}
The coefficients in the expansion of $b'(u) = \tr B(u)^2$ are mutually commuting, or equivalently
\be \left[ b'(u), b'(v) \right] = 0\ee
for all values
of $u$ and $v$. Moreover, they commute with the generators of $\liealg{gl}_N[z]^\sigma$ of degree zero:
\be [B^{(0)},b'(u)]=0\ . \ee
The algebraic elements in $B^{(0)}$ generate $\liealg{gl}_{N_0} \oplus
\liealg{gl}_{N_1} \oplus\dots\oplus \liealg{gl}_{N_{n-1}}$.
\end{proposition}
\noindent\textbf{Proof.} The details of the proof are given in
appendix \ref{AppA}.\finproof

In particular, we recover (for $n=1$) the fact that $\tr T(u)^2$ commute and (for $n=2$)
the results of Hikami \cite{hik} concerning the classical limit of the reflection algebra.

In sections \ref{gaudin} and \ref{calogero}, we will apply this
purely algebraic result to find new integrable models.

\subsection{Outer Automorphisms}

In the previous section, we focused on inner automorphisms. Now,
we show how to modify the construction to study outer
automorphisms. Modulo inner automorphisms, the only outer automorphism of
$\liealg{gl}_N$ is generalized transposition, which has order 2.

Let $K$ be a real invertible $N\times N$ matrix satisfying $K^t=\eta K$ with $\eta=\pm 1$ (for $\eta=-1$, $N$ must be even), and define an outer automorphism $\cT$ by $e_{ij} \mapsto K_{jp}
(K^{-1})_{qi} e_{pq}$, or equivalently
\be \cT: \PP_{a1} \mapsto \PP_{a1}^{\, \cT_a} = K_a \PP_{a1}^{\,t_a} K_a^{-1}=:\QQ_{a1}\ ,\label{actt}\ee
where $t_a$ is matrix transposition in the space $a$.
The eigenvalues of $\cT$ are $\pm 1$ and, as before, the decomposition of $\liealg{gl}_N$
into the direct sum of eigenspaces of $\cT$ defines a $\Z_2$-gradation.

One may introduce the $N \times N$ matrices
\begin{equation}
\label{ge} {\cal
G}^+=\mbox{diag}(\underbrace{1,\dots,1}_p,\underbrace{-1,\dots,-1}_q)~~~\mbox{and}~~~~
{\cal G}^-=\mbox{diag}(\underbrace{1,\dots,1}_{N/2})\otimes
\left(\begin{array}{cc}0& 1\\-1&0\end{array}\right)\ ,
\end{equation}
where $p+q=N$ and the second case is valid only for $N$ even. A
well-known result in linear algebra is then that $K$ is congruent
over the reals to $G^\eta$, i.e. ${\cal U} K {\cal U}^t={\cal
G}^\eta$ for some real matrix $\cal U$. From this one sees that the
$+1$-eigenspace of $\cT$ is the Lie subalgebra $\liealg{so}(p,q)$
for $\eta=+1$ and $\liealg{sp}(N)$ for $\eta=-1$.

Once more we may now define \be \liealg{gl}_N[z]^{\cT} = \left\{ A(z)\in
\liealg{gl}_N[z]\ \big|\ \cT A(z) = A(-z)\right\},\ee
that is, the subalgebra of $\liealg{gl}_n[z]$ in which each element of degree $k$ is also in the $(-1)^k$-eigenspace
of $\cT$. The projection map $\liealg{gl}_N[z]\rightarrow\liealg{gl}_N[z]^{\cT}$ is $e
z^k \mapsto \frac{1}{2}(1+(-1)^k {\cT}) e z^k$, and, given (\ref{actt}), this sends
\be T(u) \mapsto  T(u) + T(-u)^{\cT}=:S(u),\label{But}\ee
which defines the formal series $S(u)$, whose expansion in inverse powers of $u$
\be S(u)= \frac{1}{u} S^{(0)} + \frac{1}{u^2} S^{(1)} + \dots,\ee
contains a complete set of generators of $\liealg{gl}_N[z]^{\cT}$. The commutation relations of this subalgebra can be written simply by using the notation with the formal
series.

\begin{lemma} $S(u)$ obeys
\be\left[ S_a(u), S_b(v) \right] =
  \left[ S_a(u) + S_b(v), \frac{P_{ab}}{u-v}\right]+
  \left[ S_a(u) - S_b(v), \frac{Q_{ab}}{u+v}\right]\label{Bbractt}\ee
  where $Q_{ab}=P_{ab}^{{\cT}_a}=P_{ab}^{{\cT}_b}$
and has the symmetry property that \be S(u) = S(-u)^\cT
\label{unitarityt}.\ee
\end{lemma}
\noindent\textbf{Proof.} The first of these is true by virtue of
(\ref{halfloop}) and the second is immediate from the definition
(\ref{But}).\finproof

Note that these commutation relations can be obtained from the
classical limit of the twisted Yangian introduced in \cite{Ytwist}.
More abstractly, the relations (\ref{Bbractt}) and
(\ref{unitarityt}) can be regarded as defining an algebra, which can
then be seen to be embedded in the half loop algebra according to
(\ref{But}).

It is well-known that the centre of this subalgebra is generated by
the odd coefficients of the series $s(u)=\tr S(u)$ (see for example \cite{MNO}, section 4). But we have also

\begin{proposition}\label{s's'0}
The quantities in the expansion of $s'(u) = \tr S(u)^2$ are mutually
commuting, or equivalently \be \left[ s'(u), s'(v) \right] = 0\ee
for all values of $u$ and $v$. Moreover, \be [S^{(0)},s'(u)]=0\ .\ee
The elements in $S^{(0)}$ generate $\liealg{so}(p,q)$
for $\eta=+1$ and $\liealg{sp}(N)$ for $\eta=-1$.
\end{proposition}
\noindent\textbf{Proof.} The details of the proof are given in
appendix \ref{AppB}.\finproof

\section{Gaudin models}\label{gaudin}

\subsection{The Inner-twisted Gaudin Magnets \label{ing}}

The quantum Gaudin magnet, introduced in \cite{gaudin}, is an integrable spin chain with long
range interactions. The Gaudin Hamiltonians for
the model with $L$ sites are \be\cH_k=\sum_{\underset{j \neq
k}{j=1}}^L\frac{P_{jk}}{z_j-z_k}\ \ee where $z_i$ are complex numbers.
(Recall that $P_{jk}$  permutes the $j^{th}$ and
$k^{th}$ spins.) This model is usually called the $A_L$-type Gaudin
model. It may be obtained from the more general class of
integrable Hamiltonians \be\label{H1}
H_k=\sum_{\underset{j \neq k}{j=1}}^L\frac{\tr_a
\PP_{ak}\PP_{aj}}{z_k-z_j}\ee
by specifying that the spin at each site $j$ is in the fundamental
representation of $\liealg{gl}_N$.

Now, given proposition \ref{b'b'0} above, we can obtain new integrable models, as in the following proposition. These models describe spins placed at fixed positions in the plane, each of which interacts with the central point and with the other spins, not only directly, but also via their images under the rotation group of order $n$. 

 \begin{proposition}
\label{gaudintwin}
The model described by any one of the Hamiltonians \be\label{Hgin}
H_k^{(n)}=\sum_{\underset{j \neq
k}{j=1}}^L\sum_{p\in\Z_n}\frac{\tr_a\PP_{ak}G_a^{-p}\PP_{aj}G_a^{p}}{z_k-\tau^{p}z_j}
\ +\sum_{p\in\Z_n,p\neq
0}\frac{\tr_a\PP_{ak}G_a^{-p}\PP_{ak}G_a^{p}}{2z_k}\  \ee is
integrable. This model has $\liealg{gl}_{N_0} \oplus
\liealg{gl}_{N_1} \oplus\dots\oplus \liealg{gl}_{N_{n-1}}$ symmetry.
\end{proposition}

\noindent\textbf{Proof:}
 From the definition
 (\ref{Bu}) of $B(u)$, one finds
 \be b'(u)=\tr B(u)^2=\sum_{k=1}^L \sum_{j\in\Z_n}
 \frac{\tau^j}{u-\tau^{-j}z_k} H_k^{(n)} +\sum_{k=1}^L
 \sum_{j\in\Z_n}\frac{\tr_a \PP_{ak}\PP_{ak}}{(u-\tau^{-j}z_k)^2},
 \ee
with $H^{(n)}_k$ as given in the proposition. (The identity
 \be\frac{1}{(u-\tau^{-j}z_l)(u-\tau^{-k}z_p)}= \frac{1}{\tau^{-j}z_l-\tau^{-k}z_p} \left(\frac{1}{u-\tau^{-j}z_l}-\frac{1}{u-\tau^{-k}z_p}\right) \ee
 for $(j,l)\neq(k,p)$ is helpful in showing this.)

It then follows from proposition \ref{b'b'0} that $[H_k^{(n)},H_p^{(n)}]=0$. Since (for $n>1$) these operators $H_p^{(n)}$ are independent we have found $L$ commuting conserved quantities, completing the proof of integrability of the Hamiltonian $H^{(n)}$. Next, from the relation $[B^{(0)},\tr B(u)^2]=0$, also proved in proposition \ref{b'b'0}, we deduce that $[B^{(0)},H^{(n)}]=0$, which gives the $\liealg{gl}_{N_0} \oplus \liealg{gl}_{N_1} \oplus\dots\oplus \liealg{gl}_{N_{n-1}}$ symmetry of the model.
\finproof\\

 \textbf{Examples:}
 \begin{itemize}
  \item For $n=2$, we obtain the Hamiltonian
 \be \label{HBC} H^{(2)}_k=\sum_{\underset{j \neq
k}{j=1}}^L\left(\frac{\tr_a\PP_{ak}\PP_{aj}}{z_k-z_j}+
\frac{\tr_a\PP_{ak}G_a\PP_{aj}G_a}{z_k+z_j}\right)
+\frac{\tr_a\PP_{ak}G_a\PP_{ak}G_a}{2z_k}\ee of the BC-type Gaudin model studied in \cite{hik}.
 \item If the sites carry the fundamental representation of $\liealg{gl}_N$, our Hamiltonian is
 \be
H_k^{(n)}=\sum_{\underset{j \neq
k}{j=1}}^L\sum_{p\in\Z_n}\frac{G_j^{p}P_{kj}G_j^{-p}}{z_k-\tau^{p}z_j}
+\sum_{p\in\Z_n,p\neq 0}\frac{G_k^{-p}\tr G^{p}}{2z_k}\ .\ee
 \end{itemize}

Let us remark that in the $A_L$ case ($n=1$) supplementary conserved
quantities, called higher Gaudin Hamiltonians, can be found by
computing for example $\tr T(u)^3$ (see e.g. \cite{cher}). The question of whether this is possible in
the generalized cases ($n\neq 1$) studied here remains open.

\subsection{The Outer-twisted Gaudin Magnets \label{outg}}

Using the algebraic result of proposition \ref{s's'0}, we can also succeed in constructing integrable models based on outer automorphisms, as follows:

 \begin{proposition}
\label{gaudintwout}
The model described by any one of the Hamiltonians \be \label{Ht}
H_k^{\eta}=\sum_{\underset{j \neq
k}{j=1}}^L\left(\frac{\tr_a\PP_{ak}\PP_{aj}}{z_k-z_j}+
\frac{\tr_a\PP_{ak}\QQ_{aj}}{z_k+z_j}\right)
+\frac{\tr_a\PP_{ak}\QQ_{ak}}{2z_k}\  \ee is integrable. The model
has $\liealg{so}(p,q)$ symmetry (resp. $\liealg{sp}(N)$ symmetry)
for $\eta=+1$ (resp. $\eta=-1$).
\end{proposition}

\noindent\textbf{Proof.} The proof is similar to the one of
proposition \ref{b'b'0}.
 Using definition
 (\ref{But}) of $S(u)$, we show that
 \be s'(u)=\tr S(u)^2=
 \sum_{k=1}^L
 \frac{4z_k}{(u-z_k)(u+z_k)} H_k^{\eta} +\sum_{k=1}^L
\tr_a \PP_{ak}\PP_{ak}
\left(\frac{1}{(u-z_k)^2}+\frac{1}{(u+z_k)^2}\right)
 \ee
 with $H^{\eta}_k$ given as in the proposition.
 Then, we deduce from proposition \ref{s's'0} that
 $[H_k^{\eta},H_p^{\eta}]=0$, and since
 the operators $H_p^{\eta}$ are independent for different $p$, this proves the
 integrability of  $H^{\eta}$. The symmetry algebra is deduced from $[S^{(0)},\tr
 S(u)^2]=0$ proved in the proposition \ref{s's'0}.
\finproof\\

Every choice of representation  $V_1\otimes \dots \otimes V_L$ for the sites then yields a Gaudin-type model. (It is worth remarking that it is possible to choose different representations at different sites.)
For example, in the fundamental representation of $\liealg{gl}_N$, the
 Hamiltonian is
 \be
H_k^{\eta}=\sum_{\underset{j \neq
k}{j=1}}^L\left(\frac{P_{kj}}{z_k-z_j}+\frac{Q_{kj}}{z_k+z_j}\right)
+\frac{\eta}{2z_k}\ .
 \ee

We may interpret $H^\eta$ as a Gaudin model with
boundary as in the BC type model (equation (\ref{HBC}), and see also
\cite{hik}). The $z_k + z_j$ term in (\ref{Ht}) corresponds to the
interaction between the $k^{th}$ spin represented in $V_k$ and the
$j^{th}$ `reflected' spin transforming in the contragredient
representation. This type of boundary
is called soliton non-preserving and has been implemented in other
integrable models \cite{boco,gand,doikou1,doi2}. The final term in
(\ref{Ht}) corresponds to the interaction between particles and the
boundary.

\section{Calogero Models}\label{calogero}

We turn now to the second class of integrable system of interest in this work, the Calogero models.
We seek to construct dynamical models of multiple particles on a star graph, whose pairwise interactions are determined by a potential of the usual Calogero type, namely $1/q^2$, where $q$ is the linear distance separating the particles in the plane of the star graph.
We will first construct models of particles of unspecified statistics; subsequently, by specifying statistics and parity, we arrive at Calogero models for particles with internal spins.   

\subsection{The $A_L$ case}

Let us first recall the Calogero model based on the root
system $A_L$ \cite{Cal}, and in particular the use of Dunkl
operators \cite{D} in demonstrating its integrability \cite{BGHP}.
Consider a quantum mechanical system of $L$ particles on the real
line. Let $q_i$ be the position operator of the $i^{th}$ particle,
and write the position-space wave function as \be
\psi(q_1,q_2,\dots, q_L).\ee Let $\P_{ij}=\P_{ji}$ be the operator
which transposes the positions of particles $i$ and $j$, \be \P_{ij}
\psi(\dots,q_i,\dots, q_j,\dots) =
\psi(\dots,q_j,\dots,q_i,\dots).\ee 
Let us denote $S_L$ the permutation group of $L$ elements and $(ij)$ the transposition of the elements $i$ and $j$. 
Each element $s\in S_L$ can be written in terms of transpositions, namely $s=(ij)\dots (kl)$.
Then, we can define $\P_s$ as the shorthand for the product  $\P_{ij}\dots
\P_{kl}$ (even though the expression of $s$ in terms of transpositions is not unique, $\P_s$ is well-defined due to the commutation relations satisfied by $\P_{ij}$). The sign of $s$,
denoted $|s|$, is the number of these transpositions modulo 2.

Let us define $L$ operators $d_i$
-- the Dunkl operators -- by \cite{Poly,BHV} \be d_i = p_i + \lambda
\sum_{j\neq i}  \frac{1}{q_i-q_j}\P_{ij}, \quad\text{where}\quad p_i
= -i\hbar \frac{\del}{\del q_i}.\ee It follows from the relations
$\P_{ij} q_j = q_i \P_{ij}$ that the Dunkl operators commute with
one another, \be \left[ d_i, d_j \right] = 0 \ee and consequently
that the quantities \be I^{(k)} = \sum_{i=1}^L d_i^k \label{Hi}\ee
form a commuting set also. The $I^{(k)}$ are algebraically
independent for $k=1,2,\dots L$, and these give $L$ commuting
conserved quantities of the model with Hamiltonian \be H= I^{(2)} =
\sum_{i=1}^L d_i^2 = \sum_{i=1}^L \left( p_i^2
 - \sum_{j\neq i}
 \frac{1}{(q_i - q_j)^2} \lambda\left(\lambda- i\hbar\P_{ij}\right)\right),\ee
which is therefore, by construction, integrable.

The next step is to consider particles with internal degrees of freedom,
which we take to be in the fundamental representation of $\liealg{gl}_N$.
The wave funtion becomes
\be
\psi(q_1,q_2,\dots, q_L | v_1,v_2,\dots,v_L).
\ee
where $v_i\in \CC^N$.
As we define operators $\P_{ij}$ which transpose the positions, we introduce operator $P_{ij}$ 
which transposes the spins
\be
P_{ij}\psi(q_1,\dots,q_L|\dots,v_i,\dots, v_j,\dots) =
\psi(q_1,\dots,q_L|\dots,v_j,\dots,v_i,\dots).\ee 
We define similarly to $\P_s$ the matrix $P_s=P_{ij}\dots P_{kl}$ for $s=(ij)\dots(kl)$ acting on the spins.

As explained before, to use the St Petersburg notation, we need supplementary spaces 
called auxiliary spaces (which are $\CC^N$ and, in this case, isomorphic to the quantum space) 
and denoted by the letters $a$, $b$,...
The conserved quantities (\ref{Hi}) then emerge in a natural way from
the matrix
\be T_a(u) = \sum_{\ell=1}^L \frac{P_{a\ell}}{u-d_\ell},\label{Tcalg}\ee
because (as one can see using $\tr_a P_{a\ell} =1$)
\be t(u) = \tr_a T_a(u) = \sum_{k=0}^\8 \frac{I^{(k)}}{u^{k+1}}.\ee

Here (\ref{Tcalg}) is nothing but a modified version of the monodromy
matrix (\ref{realL}). The parameters $z_\ell$ are replaced by the Dunkl
operators, and since the quantum spaces are chosen to be in the fundamental
representation, $\PP_{a\ell} = E_{ij}\otimes e_{ji}$ becomes the transposition
operator $P_{a\ell} = E_{ij}\otimes E_{ji}$ (for $\ell=1,\dots, L$). Now because the $d_i$ commute with
each other \emph{and} with all operations on the internal degrees of freedom,
$T(u)$ obeys the half loop algebra relations (\ref{halfloop}) exactly as before.

Suppose, finally, that the particles are in fact indistinguishable,
which is often the case of real physical interest. One must then
impose definite exchange statistics on the wavefunction: \be P_{ij}
\P_{ij} \psi = \eps \psi,\ee where $\eps=+1$ for bosons and
$\eps=-1$ for fermions. The projector onto such states
is
\be
\Lambda = \sum_{s\in S_L} \eps^{|s|} \P_s P_s\;. \ee 

The following
relation \be (\Lambda-1) T(u) \Lambda
=0
\label{lambdaa}
\ee 
demonstrated in \cite{BGHP}
is crucial, because it implies that the
modified generators $\widetilde T(u)=T(u)\Lambda$ preserve the
condition $\Lambda \psi=\psi$, \emph{and obey the same algebraic
relations as the original $T(u)$}. From $\widetilde T(u)$ we may
define $\widetilde t(u)= \tr_a T(u) \Lambda = t(u) \Lambda$, and
hence $\widetilde I^{(k)} = I^{(k)} \Lambda$. Using $[\widetilde
t(u),\widetilde t(v)]=0$, one obtains that the $\widetilde I^{(k)}$
are once more $L$ commuting conserved quantities of the system with
Hamiltonian \be \widetilde H= \widetilde I^{(2)} = \sum_{i=1}^L
d_i^2 \Lambda = \sum_{i=1}^L \left( p_i^2 \label{HP} - \sum_{j\neq
i} \frac{ \lambda\left(\lambda- i\hbar\eps P_{ij}\right)}{(q_i -
q_j)^2} \right)\Lambda,\ee where we are now able to replace $\P$,
which acts on particle positions, by $P$, which acts only on the
internal degrees of freedom. Moreover, since $\widetilde t(u)$
commutes with $\widetilde T(u)$, the model has a half loop symmetry
algebra.

The subtlety in all this is that the Dunkl operators themselves
do \emph{not} obey any relation analogous to (\ref{lambdaa}). There
are thus essentially three steps in this procedure to construct an
integrable Hamiltonian for a system of indistinguishable particles:
\begin{enumerate}
\item Find commuting Dunkl operators, and hence $T(u)$
\item Construct the appropriate projector $\Lambda$ onto physical states,
\item Prove the relation $(\Lambda-1) T(u) \Lambda =0$.
\end{enumerate}

\subsection{Dunkl Operators for the order $n$ inner-twisted case}

\begin{figure}[ht]
\begin{center}
\epsfig{file=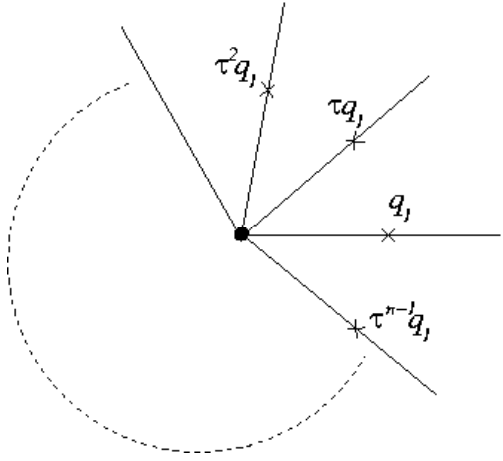,height=6cm}
\caption{A particle on the branch $\R^+$, and its images on the other branches\label{pic}}
\end{center}
\end{figure}

We can now turn to applying these ideas to the model of interest in
the present work. We consider a system of $L$ particles living on
$n$ half-lines -- ``branches'' -- joined at a central node, as in
figure \ref{pic}. The branches are given parametrically by $z=\tau^k
t, t>0$, $k\in \Zn$, and we shall denote them by \be\R^+, \tau \R^+,
\dots, \tau^{n-1} \R^+.\ee
As before, let $q_i$ be the position
operator of $i$th particle. (Note that the spectrum of $q_i$ is not real, but
only for the superficial reason that we choose to regard the half-lines as subsets of the
complex plane.) In addition to the
$\P_{ij}$, which exchange particle positions, we can define now new
operators $\Q_i$ which move the particles between branches: \be \Q_i
\psi(\dots, q_i, \dots) = \psi(\dots, \tau q_i,\dots ).\ee It is
useful to collect together the algebraic relations satisfied by the
$q_i$, $\Q_i$, and $\P_{ij}$: \be \P_{ij} \P_{jk} \P_{ij} = \P_{jk}
\P_{ij} \P_{jk}, \quad \P_{ij}^2 = 1,\quad \P_{ij}=\P_{ji},\ee \be
\Q_i^n = 1,\ee and \be \P_{ij} \Q_j = \Q_i \P_{ij}, \quad \P_{ij}
q_j = q_i \P_{ij},\quad \tau \Q_i q_i = q_i \Q_i,\label{rln2}\ee with
all the rest commuting. To construct an integrable model, the first
task is to find a suitable generalization of the commuting Dunkl
operators introduced above.
\begin{proposition} The Dunkl operators defined by
\be 
\label{eq:dunklstar}
d_i = p_i + \lambda \sum_{\underset{j\neq i}{j=1}}^L \sum_{k\in \Zn} \frac{1}{q_i-\tau^k q_j} \Q_i^k
\P_{ij} \Q_i^{-k} + \sum_{k\in \Zn}   \frac{\mu_k}{q_i}\Q_i^k, \qquad p_i = -i\hbar
\frac{\del}{\del q_i}, 
\ee 
for arbitrary parameters $\lambda, \mu_k \in \mathbb C$, commute amongst
themselves:
\be \left[d_i,d_j\right] =0.\ee 
\end{proposition}

\proof Consider first the terms at order $\lambda$.
We have \be \left[ \Q_i^k \P_{ij} \Q_i^{-k}, p_j\right] = \P_{ij}
(\tau^k p_j - p_i) \Q_j^k \Q_i^{-k} \ee using the relations
(\ref{rln2}) and the definition $p_j= -i\hbar \frac{\del}{\del
q_j}$, which together imply $\Q_j p_j = p_j \Q_j \tau$. The two terms
of this type occurring in $[d_i,d_j]$ are \bea&& \sum_{k\in\Zn}
\frac{1}{q_i-\tau^k q_j}\left[ \Q_i^k \P_{ij} \Q_i^{-k}, p_j\right]
+\frac{1}{q_j-\tau^k q_i}\left[p_i, \Q_j^k \P_{ji} \Q_j^{-k}\right]  \nn,\\
    &=&   \sum_{k\in\Zn}  \frac{1}{q_i-\tau^k q_j} \P_{ij} (\tau^k p_j - p_i) \Q_j^k \Q_i^{-k}
    - \frac{1}{q_j-\tau^kq_i}\P_{ij}(\tau^kp_i-p_j) \Q_i^k\Q_j^{-k} \eea
which cancel, after a change of the summation index in the second.
The two terms containing $[p_j, 1/(q_i-q_j)]$ cancel similarly. The
terms occurring at order $\lambda^2$ are of the form \be \left[
\frac{1}{q_i-\tau^kq_g} \Q_i^k \P_{ig} \Q_i^{-k},\frac{1}{q_j-\tau^\ell q_h} \Q_j^\ell \P_{jh}
\Q_j^{-\ell} \right].\ee These vanish
trivially unless at least one of the indices $i,g$ matches at least
one of $j,h$. It is straightforward, though tedious, to check that
the terms with exactly one index in common sum to zero, by using the
relations (\ref{rln2}) to bring every such term into e.g. the form $
\frac{1}{q-q}\frac{1}{q-q} \P\P\Q\Q\Q $ and then summing the
fractions directly. The terms in which \emph{both} indices match
give \be \sum_{k\in\Zn} \Q_i^k \Q_j^{-k} \sum_{\ell\in\Zn} \left(
\frac{1}{\tau^{-\ell} q_j - \tau^k q_i}\frac{1}{q_j-\tau^\ell q_i}
       - \frac{1}{\tau^{k-\ell}q_i-\tau^{-k} q_j}\frac{1}{q_i-\tau^{\ell-k} q_j}\right),\ee
and here the sum over $\ell$ may be re-written as
\be \frac{1}{q_i-\tau^kq_i}\sum_{\ell\in\Zn} \left( \frac{1}{q_j-\tau^\ell q_i} - \frac{1}{q_j-\tau^{k+\ell} q_i}- \frac{1}{\tau^{-k} q_j-\tau^\ell q_i} + \frac{1}{\tau^{-k} q_j-\tau^{k+\ell} q_i}\right)\ee
which then vanishes by shifting the dummy index $\ell$ in the second and fourth terms. The terms involving $\mu_k$ may be treated similarly.\finproof

These Dunkl operators have been introduced previously in \cite{duop} as Dunkl operators associated to complex reflection groups. A proof of their commutativity is already given but is based on different computations.

As in the $A_L$ case above, the quantities \be I^{(k)} =
\sum_{i=1}^L d_i^k \ee are then mutually commuting, forming a
hierarchy of Hamiltonians of an integrable system. Their detailed
forms are rather complicated -- for example, in the case of $n=3$
branches with only $L=2$ particles and $\mu_k=0$, we find that the
first three are \bea I^{(1)}&=& p_1 + p_2 + \lambda
\left(\frac{1-\tau}{q_1-\tau q_2}  \Q_1 \Q_2^{-1}  \label{n2l3}
                                         +\frac{1-\tau^2}{q_1-\tau^2 q_2} \Q_1^{-1} \Q_2 \right)\P_{12} \eea
\bea I^{(2)} &=& p_1^2 + p_2^2 \\ &&{}+ \lambda p_1 \left( \frac{1-\tau^2}{q_1-\tau q_2}\Q_1 \Q_2^{-1}  +  \frac{1-\tau}{ q_1 - \tau^2 q_2}\Q_1^{-1} \Q_2\right)\P_{12} \nn\\
 && {} + \lambda p_2 \left(\frac{1-\tau^2}{q_2-\tau q_1} \Q_2 \Q_1^{-1}  + \frac{1-\tau}{ q_2 - \tau^2 q_1}\Q_2^{-1} \Q_1 \right)\P_{12} \nn\\
 && {} - 3 \lambda^2 \frac{q_1^4 + 2q_1^3 q_2 + 2q_1 q_2^3 + q_2^4}{\left(q_1^3-q_2^3\right)^2}\nn\eea
\bea I^{(3)} &=& p_1^3 + p_2^3 \\ &&{} - 3\lambda p_1 \left( \frac{1}{(q_1-q_2)^2}(\P_{12} + \lambda) + \frac{1}{(q_1-\tau q_2)^2}(\P_{12} \Q_1^{-1} \Q_2+ \lambda)  +\frac{1}{(q_1-\tau^2 q_2)^2} (\P_{12}\Q_1 \Q_2^{-1}+\lambda) \right)\nn\\
&&{}-3\lambda p_2 \left(\frac{1}{(q_1-q_2)^2} (\P_{12} + \lambda) + \frac{1}{(q_2-\tau q_1)^2}(\P_{12} \Q_2^{-1} \Q_1+ \lambda) + \frac{1}{(q_2-\tau^2 q_1)^2}(\P_{12} \Q_2 \Q_1^{-1}+\lambda) \right)\nn\\
&&{} - 3\sqrt{3} i \lambda^2 \frac{1}{(q_1-q_2)(q_1-\tau q_2)(q_1-\tau^2 q_2)}\Q_1 \Q_2(\Q_1 - \Q_2) \nn\eea

\subsection{Quasi-partity and particles with spin}

The next steps are to consider particles with spin and to choose a suitable projector onto physical states. We take the latter to be the product of two parts
\be \Lambda_P \Lambda_Q,\ee
where
\be \Lambda_P = \sum_{s\in S_L} \eps^{|s|} \P_s P_s\ee
is the projector onto states of definite exchange statistics, and
\be \Lambda_Q = \prod_{i=1}^L \frac{1}{n} \sum_{j\in \Zn} \Q_i^j G_i^{-j}\ee
is a new projector which relates the wavefunction on different branches: $G_i\psi = \Q_i \psi$. In the case of $n=2$ this means relating the wavefunction at $q_i=x$ with $q_i=-x$, or, in other words, imposing a parity condition, so we shall call analogous conditions for arbitrary $n$ ``quasi-parity'' conditions. The reason for requiring quasi-parity is that we would like to find a Hamiltonian that does not involve $\P_{ij}$ or $\Q_j$, and, just as imposing definite statistics allowed $\P$ to be replaced by $P$ in (\ref{HP}), so here quasi-parity will allow $\Q$ to be replaced by $G$, the matrix defining the automorphism of $\liealg{gl}_N$.\footnote{Note that in fact there is another natural class of quasi-parity condition: when $n=2$, $\psi(\dots,q_i,\dots) = G_i\psi(\dots,-q_i,\dots)$ is obviously equivalent to $\psi(\dots,q_i,\dots) - G_i\psi( \dots,-q_i,\dots)=0$, but these two formulations suggest different generalisations to $n>2$: we can on the one hand demand for all $i$ that
\be  \psi(\dots,q_i,\dots) = G_i\psi(\dots,\tau q_i,\dots)=\dots = G_i^{n-1} \psi(\dots,\tau^{n-1}q_i,\dots)\nn\ee
or, alternatively, for all $i$
\be \psi(\dots,q_i,\dots) + \tau G_i \psi(\dots,\tau q_i,\dots) + \dots + \tau^{n-1} G_i^{n-1}\psi(\dots,\tau^{n-1}q_i,\dots)=0.\nn\ee
We use the first type of quasi-parity here. With such a condition in force one need only give the wavefunction on $\R^+$ in order to completely specify the state, and in this sense the model is really on the half-line. The second type of quasi-parity is weaker -- and so potentially interesting -- but does not allow us to replace $\Q$ by $G$ in the Hamiltonian.}

Now $T_a(u)$, defined in (\ref{Tcalg}), does not respect quasiparity, but we have instead
\begin{lemma} \be B_a(u) = \sum_{j\in\Zn} \tau^j G_a^j T_a(u\tau^j) G_a^{-j} \ee
satisfies
\be \left[ B_a(u), \Lambda_Q \right] = 0.\ee
\end{lemma}
\proof By direct computation. \finproof

Thus, using also that $T(u)$ obeys (\ref{lambdaa}), we have
\be (1 - \Lambda_P \Lambda_Q) B_a(u) \Lambda_P \Lambda_Q = 0.\ee
and have therefore arrived at the following result

\begin{proposition}
\label{caltw}
The modified generators $\widetilde B(u) = B(u)\Lambda_P \Lambda_Q$ preserve the statistics and quasiparity of the wave function and themselves satisfy the relations (\ref{Bbrac}).
The quantities $\widetilde I^{(k)}$ in the expansion of
\be\widetilde b(u) = \tr_a B_a(u)\Lambda_P \Lambda_Q =: \sum_{k=0}^\8 \frac{\widetilde I^{(k)}}{u^{k+1}}\ee
are mutually commuting, and non-zero only when $k\equiv 0\mod n$. The model with this hierarchy of integrable Hamiltonians has symmetry $\liealg{gl}_N[z]^\sigma$.
\end{proposition}
\proof
Most of this follows from the construction above: it remains only to show that $\widetilde I^{(k)}$ vanishes for $k\not\equiv 0 \mod n$. One sees this by writing
\be\widetilde b(u) = \sum_{\ell=1}^L \sum_{p=0}^\8 \left( \sum_{k\in\Zn} \tau^{-pk} \right) \frac{d^{\,p}_l}{u^{p+1}}\Lambda_P \Lambda_Q \ee
and noting that $\sum_{k\in \Zn} \tau^{-pk}$ is zero unless $p\equiv 0\mod n$. \finproof

The vanishing of some of the generators is as expected: in the case $n=2$, for example, the charge $I^{(1)}$, which is first order in momentum $p$, does not survive the introduction of a boundary. In the case of $n=3$ with $L=2$ particles, already mentioned in
(\ref{n2l3}), the first non-vanishing charge is third order in momentum:
\bea \widetilde I^{(3)} &=& p_1^3 + p_2^3 \\ &&{} - 3\lambda p_1 \left( \frac{1}{(q_1-q_2)^2}(\eps P_{12} + \lambda) + \frac{1}{(q_1-\tau q_2)^2}(\eps P_{12} G_1^{-1} G_2+ \lambda)  +\frac{1}{(q_1-\tau^2 q_2)^2} (\eps P_{12}G_1 G_2^{-1}+\lambda) \right)\nn\\
&&{}-3\lambda p_2 \left(\frac{1}{(q_1-q_2)^2} (\eps P_{12} + \lambda) + \frac{1}{(q_2-\tau q_1)^2}(\eps P_{12} G_2^{-1} G_1+ \lambda) + \frac{1}{(q_2-\tau^2 q_1)^2}(\eps P_{12} G_2 G_1^{-1}+\lambda) \right)\nn\\
&&{} - 3\sqrt{3} i \lambda^2 \frac{1}{(q_1-q_2)(q_1-\tau q_2)(q_1-\tau^2 q_2)}G_1 G_2(G_1 - G_2)\ . \nn\eea

\section{Conclusion}

In this paper, we gave the St Petersburg presentation of subalgebras
of the $\liealg{gl}_N$ half loop algebras associated to all finite
order automorphisms of $\liealg{gl}_N$. This presentation allows us
to obtain commuting quantities used to prove integrability for new
integrable models of Gaudin or Calogero type. The non-Abelian
symmetry for each of these new models is also exhibited.

We may expect that the usual methods to solve the Gaudin models and the Calogero models may be generalized to solve the models
given in the propositions \ref{gaudintwin}, \ref{gaudintwout} or \ref{caltw} introduced in this paper.
Namely, for the Gaudin models, the Hamiltonians (\ref{Hgin}) and (\ref{Ht}) may be diagonalised by generalising the usual methods such
as the separation of variables \cite{skly_sov} or the algebraic Bethe ansatz \cite{fad}.
For the Calogero models, the previous link established in \cite{duop} between the nonsymmetric Jack polynomials and the Dunkl operators
(\ref{eq:dunklstar}) may be useful to diagonalize the Hamiltonian given in proposition \ref{caltw}.

Our discussion has dealt exclusively with quantum mechanical models. However, for each algebra
introduced in the paper, there exists an associated Poisson bracket
algebra, obtained simply by replacing the commutator on the left of
the defining relations (\ref{HNalg}), (\ref{lie}), (\ref{halfloop}),
(\ref{Bbrac}) and (\ref{Bbractt}) by a Poisson bracket. This allows us
to treat certain classical mechanical problems. In such problems the
entries of $T(u)$, $B(u)$ or $S(u)$ are commuting functions on phase
space, which simplifies many computations. For example, the results
of propositions \ref{b'b'0} and \ref{s's'0} are replaced by the stronger statements
\be \left\{ b_k(u), b_\ell(v) \right\} =
0\quad\mbox{and}\quad \left\{ s_k(u), s_\ell(v) \right\} =
0\label{compbS}\ee where $b_k(u) = \tr B(u)^k$ and $s_k(u) = \tr
S(u)^k$. These results strongly suggest that the classical
counterpart of models given by (\ref{Hgin}) and (\ref{Ht}) are
integrable in the sense of Liouville. (It remains to prove that the quantities are independent.)

Finally, although the models of this paper were quantum-mechanical, the \emph{algebras} are classical, in the sense that they are not $q$-deformed. A very interesting question is whether a similar construction of subalgebras from higher order automorphisms of $\liealg{gl}_N$ is possible in the case of quantum groups. If so, then these subalgebras would be a $q$-deformation of those in this paper, and should also have associated to them integrable models on $n$ half-lines.

\textbf{Acknowledgements:} NC is grateful for the
financial support of the TMR Network "EUCLID. Integrable models and
applications: from strings to condensed matter", contract number
HPRN-CT-2002-00325. CASY gratefully acknowledges the financial
support of PPARC.

\appendix
\section{Proof of proposition \ref{b'b'0}} \label{AppA}
It is convenient first to use
\be G_a^{-k} P_{ab} G_a^k \,B_a(u) = G_a^{-k} G_b^k \,B_b(u) \,P_{ab} = G_b^k B_b(u) (G_b^{-k} G_b^k) G_a^{-k} P_{ab}= G_b^k B_b(u) G_b^{-k}\, G_a^{-k} P_{ab} G_a^k\nn,\ee
and similar manipulations, to re-write the commutation relations (\ref{Bbrac}) as
\bea \left[ B_a(u), B_b(v) \right] &=& \sum_{k\in \Zn}  \label{Bbrac2}
      \left(B_b(v) + \tau^k B_a(u) - \tau^k B_a(\tau^k v) - B_b(\tau^{-k}u) \right) \frac{G_a^{-k} P_{ab} G_a^k}{u-\tau^kv}.\eea
The goal here, and in the following, is to bring every term containing $P_{ab}$ into the form $B_a B_b \,G_a^{-k}P_{ab}G_a^k$. Next, we have
\bea \left[ B_a(u), B_b(v)^2 \right] &=&   B_b(v) \left[ B_a(u), B_b(v) \right] +\left[ B_a(u), B_b(v) \right] B_b(v)\\
             &=& \sum_{k\in\Zn} \Big\{ B_b(v) \left( B_b(v) + \tau^k B_a(u) - \tau^k B_a(\tau^k v) - B_b(\tau^{-k} u)\right) \nn\\ &&\qquad{}+ \left( B_b(v) + \tau^k B_a(u) - \tau^k B_a(\tau^k v) - B_b(\tau^{-k} u)\right) \tau^k B_a(\tau^k v) \Big\}
                \frac{G_a^{-k} P_{ab} G_a^k}{u-\tau^kv}\nn\\
&=& \sum_{k\in\Zn} \Big\{ B_b(v)^2 - B_b(v) B_b(\tau^{-k} u) +
                          \tau^{2k} B_a(u) B_a(\tau^k v) -\tau^{2k}B_a(\tau^k v)^2\nn\\
&& \qquad{} - \tau^k B_a(\tau^kv) B_b(\tau^{-k}u) + \tau^k B_a(u)B_b(v) \nn\\
&& \qquad{} + \tau^k [B_a(\tau^kv), B_b(\tau^{-k}u)] - \tau^k [B_a(u),B_b(v)]\Big\} \frac{G_a^{-k} P_{ab} G_a^k}{u-\tau^kv} .\nn\eea
and then the brackets in the final line may be evaluated by using (\ref{Bbrac2}) once more, to give, after some manipulation of the summation indices,
\bea \left[ B_a(u), B_b(v)^2 \right] &=& \sum_{k\in\Zn} \Big\{ B_b(v)^2 - B_b(v) B_b(\tau^{-k} u) +
                                                              \tau^{2k} B_a(u) B_a(\tau^k v) - \tau^{2k}B_a(\tau^k v)^2\nn\\
&& \quad\qquad{} - \tau^k B_a(\tau^kv) B_b(\tau^{-k}u)
                 + \tau^k B_a(u)B_b(v) \Big\} \frac{G_a^{-k} P_{ab} G_a^k}{u-\tau^kv}\label{bb2}\\
&&- \sum_{j,k\in\Zn} \Big\{  \tau^{2j} B_a(\tau^jv) - \tau^{2j} B_a(\tau^{j-k}u) - \tau^{k} B_b(\tau^{k-j} v) \nn\\
&&  \quad\qquad{} + \tau^k B_b(v) + \tau^{j+k} B_a(u) - \tau^{j+k} B_a(\tau^j v) \Big\}
                       \frac{G_a^{k-j} G_b^{j-k}}{(u-\tau^k v)(u-\tau^j v)} \nn.\eea
Therefore
\bea \left[ B_a(u)^2, B_b(v)^2\right] &=& B_a(u) \left[ B_a(u), B_b(v)^2\right] +\left[B_a(u),B_b(v)^2\right] B_a(u)\label{b2b2}\\
&=& \sum_{k\in\Zn} B_a(u) \Big\{ B_b(v)^2 - B_b(v) B_b(\tau^{-k} u) +
                                                 \tau^{2k} B_a(u) B_a(\tau^k v) - \tau^{2k}B_a(\tau^k v)^2\nn\\
&& \quad\qquad{} - \tau^k B_a(\tau^kv) B_b(\tau^{-k}u) + \tau^k B_a(u)B_b(v) \Big\} \frac{G_a^{-k} P_{ab} G_a^k}{u-\tau^kv}\nn\\
&& + \sum_{k\in\Zn} \Big\{ B_b(v)^2 - B_b(v) B_b(\tau^{-k} u) +
                                                              \tau^{2k} B_a(u) B_a(\tau^k v) - \tau^{2k}B_a(\tau^k v)^2\nn\\
&& \quad\qquad{} - \tau^k B_a(\tau^kv) B_b(\tau^{-k}u) + \tau^k B_a(u)B_b(v) \Big\}
                                          \tau^{-k} B_b(\tau^{-k} u) \frac{G_a^{-k} P_{ab} G_a^k}{u-\tau^kv} \nn\\
&& - \sum_{j,k\in\Zn} B_a(u) \Big\{ \tau^{2j} B_a(\tau^jv) - \tau^{2j} B_a(\tau^{j-k}u) - \tau^{k} B_b(\tau^{k-j} v) \nn\\
&&  \quad\qquad{} + \tau^k B_b(v) + \tau^{j+k} B_a(u) - \tau^{j+k} B_a(\tau^j v)\Big\}
                       \frac{G_a^{k-j} G_b^{j-k}}{(u-\tau^k v)(u-\tau^j v)} \nn\\
&& - \sum_{j,k\in\Zn} \Big\{ \tau^{2j} B_a(\tau^jv) - \tau^{2j} B_a(\tau^{j-k}u) - \tau^{k} B_b(\tau^{k-j} v) \nn\\
&&  \quad\qquad{} + \tau^k B_b(v) + \tau^{j+k} B_a(u) - \tau^{j+k} B_a(\tau^j v) \Big\}
                       \frac{G_a^{k-j} G_b^{j-k}}{(u-\tau^k v)(u-\tau^j v)} B_a(u)\nn\eea
In the final line we could again add commutators to move all the $B_a$'s to the left of the $B_b$'s, but in fact it is not necessary to do so in order to evaluate
\be \left[ b'(u), b'(v) \right] = \tr_{ab} \left[ B_a(u)^2, B_b(v)^2 \right].\ee
Consider first those terms containing $P_{ab}$. Because we avoided terms of the type $B_b B_a B_b P_{ab}$ in (\ref{b2b2}), these all reduce to single traces: for example
\bea&& \tr_{ab} B_a(u) B_b(v)^2 G_a^{-k} P_{ab} G_a^k = \tr_{ab} G_a^k B_a(u) G_a^{-k} B_b(v)^2 P_{ab}\nn\\
   &=&\tr_{ab} \tau^{-k} B_a(\tau^{-k}u) P_{ab} B_a(v)^2 = \tr\tau^{-k} B_a(\tau^{-k}u) B_a(v)^2,\nn\eea
where the final equality is valid because $\tr_b P_{ab} = 1$. The remaining terms give products of traces, and,
after some cancellation, one finds
\bea \left[ b'(u), b'(v) \right]&=&
  \sum_{k\in\Zn} \frac{2\tau^{-k}}{u-\tau^kv} \tr \left[ B(\tau^{-k} u)^2, B(v) \right]\\
&&+\sum_{j,k\in\Zn} \Big\{\tr \tau^j B(u) B(\tau^j v) G^{k-j}  \tr G^{j-k}
+ \tau^k \tr B(\tau^k v) B(u) G^{k-j} \tr G^{j-k}  \nn\\
&&\quad- \tr B(u)G^{k-j}  \tr  B(v)G^{j-k}
   -\tr  B(v)G^{j-k} \tr  B(u)G^{k-j} \Big\} \frac{\tau^k-\tau^j}{(u-\tau^k v)(u-\tau^j v)} \nn.\eea
The first term, cubic in $B$, reduces, given the identity $\tr [M,N] = \tr_{ab} [M_a, N_b] P_{ab}$ and (\ref{bb2}), to
\bea &&\sum_{j,k\in\Zn} \Big\{\tr B(\tau^j v) B(u) G^{j-k} \tr G^{k-j} - \tr B(u) B(\tau^j v) G^{k-j} \tr G^{j-k} \\
    &&\qquad- \tau^{-j} \tr B(u) G^{j-k} \tr B(v) G^{k-j} + \tau^{-j} \tr B(v) G^{j-k} \tr B(u) G^{k-j} \Big\}\frac{2\tau^{j+k}}{(u-\tau^kv)(u-\tau^jv)}\nn,\eea
and, on collecting terms, one has
\bea \left[ b'(u), b'(v) \right] &=& \sum_{j,k\in\Zn} \Big\{ \tau^j  \tr\left[ B(\tau^jv)G^{k-j}, B(u) \right]\tr G^{j-k}\\
                &&\quad\qquad+ \left[\tr B(v) G^{j-k}, \tr B(u) G^{k-j} \right]\Big\}  \frac{\tau^j + \tau^k}{(u-\tau^jv)(u-\tau^kv)}.\nn\eea
The second commutator can be shown to vanish, and on evaluating the first one is left with
\bea \left[ b'(u), b'(v) \right] &=& \sum_{j,k,l\in\Zn} \Big\{ \label{uvf}
       \tr B( \tau^{-l} u) G^{k-l} \tr G^{l-j} - \tr B(\tau^{-l} u) G^{l-j} \tr G^{k-l} \\
&&\quad\qquad+ \tr B(v) G^{l-j} \tr G^{k-l} - \tr B(v) G^{k-l} \tr G^{l-j} \Big\}
\frac{(\tau^j + \tau^k)\tr G^{j-k} }{(u-\tau^jv)(u-\tau^kv)(u-\tau^lv)}\nn.\eea
Consider now the two terms containing $B(v)$. After taking $\frac{1}{3}$ the sum over the cyclic permutations of the dummy indices $i,j,k$, one finds that these reduce to
\be \sum_{j,k,l\in\Zn} \tr B(v) G^{j-k} \tr G^{k-l}\tr G^{l-j}
                    \frac{\tau^k - \tau^j}{(u-\tau^jv)(u-\tau^kv)(u-\tau^lv)}\ee
and the coefficient of $\tr B(v)G^{-a-b} \tr G^a \tr G^b$ in this sum is (with a factor $\half$ when $a=b$)
\be \sum_{l\in\Zn} \frac{ \tau^{a+l} - \tau^{b-l} }{(u-\tau^{a+l}v)(u-\tau^{l-b}v)(u-\tau^lv)}+
             \frac{ \tau^{b+l} - \tau^{a-l} }{(u-\tau^{b+l}v)(u-\tau^{l-a}v)(u-\tau^lv)},\ee
which may be seen to vanish by using $\frac{\tau^{a+l} - \tau^{l-b}}{(u-\tau^{a+l}v)(u-\tau^{l-b}v)} = \frac{1/v}{u-\tau^{a+l} v} -\frac{1/v}{u-\tau^{l-b}}$, and the same identity with $(a\leftrightarrow b)$. Similar arguments hold for the $B(u)$ terms in
(\ref{uvf}), and we have, finally, that
\be  \left[ b'(u), b'(v) \right] =0.\ee
It remains to show that $[B^{(0)},b'(u)]=0$. This may be seen by expanding (\ref{bb2}) to leading order in $1/u$ and taking the trace in space $b$.
The elements in $B^{(0)}$ is the +1-eigenspace of $\sigma$ therefore they generate $\liealg{gl}_{N_0} \oplus
\liealg{gl}_{N_1} \oplus\dots\oplus \liealg{gl}_{N_{n-1}}$.

\section{Proof of proposition \ref{s's'0}} \label{AppB}

Let us rewrite the relation (\ref{Bbractt}) as \be\left[ S_a(u),
S_b(v) \right] =
  \left( S_a(u) + S_b(v)-S_b(u)-S_a(v)\right) \frac{P_{ab}}{u-v}+
  \left[ S_a(u) - S_b(v), \frac{Q_{ab}}{u+v}\right]\ .\label{Bbracapp}\ee
Note that, in contrast to the previous case computed in appendix \ref{AppA}, here the $Q$ cannot be
moved through the $S$. We are now in the position to compute the
bracket $\left[ S_a(u), S_b(v)^2 \right]$ by again using
(\ref{Bbracapp}) to bring every term containing $P_{ab}$ on the
right-hand side into the form $S_a S_b P_{ab}$. We find
\begin{eqnarray} \label{relS12}
\left[ S_a(u), S_b(v)^2 \right]&\!\!\!\!
=&\!\!\!\!\Big(S_a(u)\big(S_a(v)+S_b(v)\big)-S_a(v)^2-\big(S_a(v)+S_b(v)\big)S_b(u)
+S_b(v)^2\Big)\frac{P_{ab}}{u-v}\quad\quad\nonumber\\
&&\!\!\!\!+S_a(u)\frac{Q_{ab}}{u+v}S_b(v)-S_b(v)\frac{Q_{ab}}{u+v}S_a(u)
+\Big[\big(S_a(u)-S_b(v)\big)S_b(v),\frac{Q_{ab}}{u+v}\Big]\nonumber\\
&&\!\!\!\!+\Big[S_a(v)-S_b(u)-S_a(u)+S_b(v),\frac{\eta
Q_{ab}}{u^2-v^2}\Big]
\end{eqnarray}
where we used also the property $P_{ab}Q_{ab}=\eta
Q_{ab}=Q_{ab}P_{ab}$. Now we can compute $\left[ S_a(u)^2, S_b(v)^2
\right]=S_a(u)\left[ S_a(u), S_b(v)^2 \right]+\left[ S_a(u),
S_b(v)^2 \right]S_a(u)$ and take the trace in spaces $a$ and $b$.
It is then straightforward to show that \be \label{s-s}
[s'(u),s'(v)]=\frac{2}{u-v}\tr[S(u)^2,S(v)]+\frac{1}{u+v}
\Big(\tr[S(u)^2,S(-v)]-\tr[S(-u)^2,S(v)]\Big)\ ,\ee where we have
used the symmetry relation (\ref{unitarityt}) and, for example, the
following properties \begin{eqnarray}&& \tr_{ab}
S_a(u)Q_{ab}S_b(u)=\tr_{ab}
S_a(u)Q_{ab}S_a(u)^\cT=\tr S(u)S(u)^\cT\\
&&\tr_{ab} Q_{ab}S_a(u)S_b(v)S_a(u)=\tr_{ab} S_b(u)^\cT
S_b(v)S_b(u)^\cT Q_{ab}=\tr S(u)^\cT S(v)S(u)^\cT\ .
\end{eqnarray}
Next, using the property \be
\tr[S(x)^2,S(y)]=-\tr[S(y),S(x)^2]=-\tr[S_a(y),S_b(x)^2]P_{ab} \ee
and relation (\ref{relS12}), we have that
$\tr[S(x)^2,S(y)]=\frac{N}{x-y}\tr[S(y),S(x)]=\frac{N}{x-y}\tr[S_a(y),S_b(x)]P_{ab}$.
Then using relation (\ref{Bbractt}), we get $\tr[S(x)^2,S(y)]=0$
which implies that the R.H.S. of (\ref{s-s}) vanishes and proves that $[s'(u),s'(v)] =  0$.
Finally, expanding (\ref{relS12}) to first order in $1/u$ and taking the trace in space $b$ yields $[S^{(0)},tr
S^2]=0$. The elements in $S^{(0)}$ is the +1-eigenspace of $\cT$ therefore they generate $\liealg{so}(p,q)$
for $\eta=+1$ and $\liealg{sp}(N)$ for $\eta=-1$. This completes
the proof of proposition \ref{s's'0}.

\end{document}